\documentclass[12pt]{article}
\pdfoutput=1
\usepackage{graphicx}
\usepackage{color}
\usepackage{cancel}
\usepackage{comment}
\usepackage{amsmath,amssymb,booktabs,xcolor,cite}

\newcommand{\lsim}{\raisebox{-0.13cm}{~\shortstack{$<$ \\[-0.07cm] $\sim$}}~} 
 
\newcommand{\beq}{\begin{eqnarray}} 
\newcommand{\eeq}{\end{eqnarray}} 
 
\newcommand{\shat}{\hat s}

\begin{document}

\begin{flushright}
{\tt KCL-PH-TH/2026-28, CERN-TH-2026-286}
\end{flushright} 

\vspace*{5mm} 

\begin{center}

\mbox{\large\bf 
Interference effects in MSSM Higgs boson searches at low $\mathbf{tan\beta}$:} 

\vspace*{2mm}

\mbox{\large\bf the  $\mathbf{gg\to H\to WW, ZZ, hh}$ and $\mathbf{gg \to A \to Zh}$ channels} 

\vspace*{8mm}

{\sc Abdelhak~Djouadi$^{1}$} and  {\sc John Ellis$^{2,3}$} 

\vspace{6mm}

{\small 
\mbox{ \hspace*{-6mm} $1$ Departamento de F\'isica Te\'orica y del Cosmos, Universidad de Granada,
18071 Granada, Spain.} \\[2mm] 

$^2$ Theoretical Physics Department, CERN, CH 1211 Geneva, Switzerland.\\[2mm]

$^3$ Physics Department, King’s College London, Strand, London, WC2R 2LS, United Kingdom.
}

\end{center}

\vspace*{8mm}

\begin{abstract}
\noindent
The LHC limits on the heavy neutral Higgs bosons of the minimal supersymmetric extension of the Standard Model, which are particularly relevant at low values of the ratio of vacuum expectation values $\tan\beta$, have been obtained in the narrow-width approximation, without including the interference between the pure Higgs signal and the QCD background. We study the interference of the resonant heavy neutral CP-even $H$ and CP-odd $A$ boson contributions with the non-resonant gluon-fusion amplitudes in the processes $gg\to VV$ with $V=W,Z$, $gg \to hh$ and $gg\to Zh$ at low $\tan\beta$,  $1\le\tan\beta\le 6$. We work at leading order but with the exact top-quark mass dependence, and with the longitudinal $VV$ and $Zh$ amplitudes obtained from the Goldstone-boson equivalence theorem that lead to loop form factors similar to those appearing in $gg\to hh$. The lineshapes in the four search channels are governed by a simple complex parameter $\kappa$,  the ratio of the resonant to non-resonant amplitudes at the peak. Because the $H$ and $A$ bosons couple weakly to the considered final states, while being broadened by their decays to top quarks, $|\kappa|$ drops to values of a few for masses of $0.5$--$1$~TeV, even though the total widths remain at the percent level. The interference then increases the signal rates by up to tens of per cent, and distorts their lineshapes. For the present ATLAS (and presumably CMS) Run II exclusion limits, the net effect is nevertheless rather modest, except at the high-mass end of the $H\to hh$ contour, but it becomes important for the masses that will be probed at the high-luminosity LHC.
\end{abstract}

\newpage

\subsection*{1. Introduction}
\label{sec:intro}

Supersymmetry remains one of the best-motivated extensions of the Standard Model (SM) and searches for its manifestations are still high on the agendas of the LHC experiments. The minimal supersymmetric extension of the SM (MSSM) makes a definite prediction for the Higgs sector: it requires only two Higgs doublet fields which, after electroweak symmetry breaking, lead to five physical states. These are the light CP-even boson $h$, identified with the observed $125$~GeV Higgs boson \cite{LHC-125}, a heavy CP-even $H$, a CP-odd $A$ and a pair of charged $H^\pm$ bosons; see Ref.~\cite{Djouadi:2005gj} for a review. Radiative corrections play an important role in this Higgs sector, but once the measured mass of $h$ is used to absorb their dominant effects, the sector can be described, to a good approximation, by only two parameters,  one Higgs mass that is taken to be that of the $A$ state $M_A$ and the ratio $\tan\beta$ of the two vacuum expectation values. This is the hMSSM~\cite{hMSSM}, which is used by the LHC experiments to present their MSSM Higgs constraints in the $[M_A,\tan\beta]$ plane~\cite{ATLPUB2024008,CMS:2024phk}.

The region of low $\tan\beta$, $1\lesssim\tan\beta\lesssim5$, is particularly interesting~\cite{hMSSM,Bagnaschi:2018ofa}. There, the neutral heavy states are produced mainly by gluon fusion through top-quark loops, $gg\to H/A$, and, in addition to their dominant decays into top-quark pairs, they have sizable branching ratios into final states that are well measured at the LHC: $H\to ZZ$, $WW$ and $hh$, and $A\to Zh$. Since the light Higgs boson is SM-like, the couplings of $H$ to gauge bosons and of $A$ to $Zh$ pairs are in general suppressed but they are still significant at low $\tan\beta$, and these channels provide important constraints for $M_A\lesssim1$~TeV~\cite{ATLPUB2024008,ATLAS:ZZ,ATLAS:Zh,ATLAS:hh2023}.

All these final states are also produced in gluon fusion through non-resonant SM processes, $gg \to VV$ with $V=W,Z$, $gg \to Zh$ and $gg \to hh$, generated by box diagrams with virtual light and/or top quarks. As the initial- and final-state topologies are the same, the corresponding amplitudes interfere with the resonant ones. While interference effects have been discussed intensively in various channels,~\footnote{They have been mostly discussed for the production of the SM-like $h$ boson via $gg\to h$ and its decays into $\gamma\gamma$ and $Z\gamma$ final states (as well as $WW,ZZ$ that we will shortly address), and in the case of heavy $H/A$ production when the dominant decays are $A/H \to t\bar t$; see Refs.~\cite{Martin:2012xc,Hgamma,Buccioni:2023qnt,MSSM-tt} for the most recent analyses.} relatively few studies have been made for the MSSM channels listed above.  In particular, the experimental interpretations generally assume narrow resonances without interference; the latter being at most considered for large widths, $\Gamma/M\gtrsim5\%$, for Higgs states with SM-like couplings~\cite{ATLAS:2017spa,ATLAS:ZZ}. 

More analyses have been performed on the theoretical side in various scenarios. Interference between a heavy $H$ decaying into $VV$ and the $gg\to VV$ continuum has been studied for a heavy SM-like Higgs boson \cite{SM-VV} in the singlet extension of the SM~\cite{Singlet-VV,Kauer:2015hia}, and in a two-Higgs-doublet model (2HDM)~\cite{Greiner:2015ixr}. These studies established that the interferences of the heavy state with the off-shell light one and with the top-quark box continuum have opposite signs and largely cancel near the resonance~\cite{Kauer:2015hia,ATLAS:2017spa}. Lineshape distortions and their dependence on the width scheme were discussed in Ref.~\cite{Lineshape}. In the MSSM, the full one-loop $gg \! \to \! ZZ$ amplitude was computed long ago \cite{Berger:1998hc} and destructive interference was  observed just below the heavy-Higgs peak. For resonant di-Higgs production, interference with the non-resonant $gg \! \to \! hh$ amplitude has been studied in the singlet Higgs model~\cite{hh-Singlet} and in the 2HDM~\cite{hh-2HDM}, where it can change the resonant rate by several tens of per cent. For $gg \!\to \! A\to Zh$, the interference with the continuum was included in Ref.~\cite{Harlander:2013mla}. In the $t\bar t$ channel, where the continuum is tree-level QCD, interference is the dominant effect~\cite{MSSM-tt} and is included in the experimental analyses~\cite{ATLAS:2024tt}.

In this Letter we consider the four possible MSSM channels at low $\tan\beta$, namely $gg\to WW,ZZ,hh$ and $Zh$, and describe the interference effects in all of them in a common framework~\footnote{This framework was set out long ago by U. Fano \cite{Fano:1961zz} and is described in the Appendix.}. We show that the usual criterion of a large Higgs width to mass ratio $\Gamma/M$ is not appropriate in our case and that what matters is, instead, the ratio $\kappa$ of the resonant to the non-resonant amplitude,  which is small when the Higgs couplings to the final state are suppressed, as is the case in the MSSM. We compute the ratio $\kappa$ at leading order and using the Goldstone-boson equivalence theorem (which is a sufficient approximation for our purpose) and including the exact dependence on the top-quark mass in the loop diagrams (which is important as we are dealing with Higgs masses beyond the $2m_t$ threshold). We then estimate how much the experimental exclusion contours change in the relevant parameter space, i.e., $1 \leq \tan\beta \leq 6$ and $M_A \lsim 1$ TeV. 
In particular, we translate the resulting impact on the comprehensive summary of the Higgs constraints in the hMSSM given by the ATLAS collaboration~\cite{ATLPUB2024008}.

\subsection*{2. Theoretical framework}
\label{sec:framework}

In the MSSM, the CP-even Higgs mass matrix in the basis of the neutral components of the two Higgs doublets is given, in terms of the mass $M_A$ and the angle $\beta$, by
\begin{equation}
\mathcal M^2 = M_Z^2\begin{pmatrix} c_\beta^2 & -s_\beta c_\beta\\ -s_\beta c_\beta & s_\beta^2\end{pmatrix}
+M_A^2\begin{pmatrix} s_\beta^2 & -s_\beta c_\beta\\ -s_\beta c_\beta & c_\beta^2\end{pmatrix}
+\begin{pmatrix}0&0\\0&\epsilon\end{pmatrix},
\end{equation}
where $s_\beta\equiv \sin\beta, c_\beta \equiv \cos\beta $  and the diagonalization gives the CP-even Higgs masses $M_H$ and $M_h$ and their mixing angle $\alpha$, and the couplings relative to the SM ones are
\begin{equation}
g_{hVV}=s_{\beta -\alpha},\quad g_{HVV}=c_{\beta-\alpha},\quad g_{htt}={c_\alpha}/{s_\beta},\quad g_{Htt}={s_\alpha}/{s_\beta},\quad g_{Att}=\cot\beta .
\end{equation}
The quantity $\epsilon$ represents the dominant radiative correction in the Higgs sector, which turns out to be rather large \cite{Djouadi:2005gj,Slavich:2020zjv}. It can be  fixed by requiring that the $h$ mass has its measured value  $M_h=125.1$~GeV~\cite{hMSSM}. $M_H$ and $\alpha$ (as well as $M_{H^\pm}$) are then uniquely determined in terms of $\tan\beta$ and $M_A$ as at LO.  For the trilinear couplings $\lambda_{hhh}$ and $\lambda_{Hhh}$, we use the simple approximate expressions in the hMSSM~\cite{hMSSM}, which are also sufficient for our purpose here. The coupling of $Ah$ to the neutral Goldstone boson, to be used later, is  $\lambda_{AG^0h} \!=\! (M_h^2\!-\!M_A^2)\cos(\beta\!-\!\alpha)/v$, where $v\!=\sqrt{v_1^2\!+\!v_2}=246$ GeV. 

Since the interferences depend strongly on the total decay widths, the $H$ and $A$ widths are computed including the QCD corrections to the dominant decays \emph{$H,A\to t\bar t$}. We use the full NLO correction with the top-quark pole mass near threshold, and the running mass $\overline m_t(M)$ with the NNLO QCD factor $1+5.67\,\alpha_s/\pi+(35.94-1.36\,n_f)(\alpha_s/\pi)^2$ far above it \cite{Hdecays}; the two are interpolated linearly in $4m_t^2/M^2$, in the spirit of the program {\sc HDECAY}~\cite{HDECAY}. The decays \emph{$H,A\to b\bar b$} are subleading in our case, and are computed with the running $b$-quark mass and the same QCD factor as above, while the \emph{$H,A\to gg$} widths use the exact LO top-quark loop and a $K$-factor of $1.7$. The decays \emph{$H\to VV$, $hh$ and $A\to Zh$} are computed at LO; they receive no QCD corrections at this order.

In terms of the velocity $\beta_t \! =\! \sqrt{1\!-\!4m_t^2/M^2}$, 
the dominant $H,A \!\to\! t\bar t$ widths are
\beq
\Gamma_H \simeq 3G_Fm_t^2M_Hg_{Htt}^2\,\beta_t^3/(4\sqrt2\pi) \  \quad {\rm and} \quad  
\Gamma_A \simeq 3G_F m_t^2M_A g_{Att}^2\,\beta_t/(4\sqrt2\pi) \, ,
\eeq
at LO, with significant QCD corrections.  The corrected widths increase just above the $t\bar t$ threshold, through the Coulomb enhancement of the slowly-moving top quarks: by $6\%$ for $H~ (\propto 
\beta_t^3)$ and up to $90\%$ for $A~(\propto \beta_t)$ at $M_A=350$~GeV. Well above the $t\bar t$ threshold they decrease, by $6$, $12$ and $15\%$ at $M_A=600$, $800$ and $1000$~GeV, respectively. 
The couplings to the final states of interest are small, as they are proportional to $\cos(\beta-\alpha)$ which ranges between $\approx-0.07$ to $\approx - 0.016$ for $M_A=400$--$800$~GeV at $\tan\beta=2$, and the widths remain below $5\%$ of the masses for $1\le\tan\beta\le6$ and $M_A\le1$~TeV. 

We come now to the main issue of the analysis, namely the amplitudes of the various processes under consideration. We compute the gluon-fusion amplitudes at LO with exact top-quark mass dependence; $b$-quark loops are negligible at low $\tan\beta$ and squark loops are absent in the hMSSM. This LO treatment is sufficient, as we give our results as ratios to the narrow-width signal that ATLAS already uses for normalization. The QCD K-factors are similar for the two amplitudes, so they largely cancel as we are normalizing to the ATLAS case. The inclusion of the full $m_t$ dependence is instead very important, as the behavior will be very different above the $2m_t$ threshold and below it (where the heavy top mass approximation is generally sufficient).  

Only the part of the non-resonant amplitude with the same angular momentum ($J=0$) and gluon helicities as the Higgs resonance interferes with the signal amplitude. For the longitudinally-polarized $Z$ bosons, which dominate the decays of a heavy scalar state, we use the Goldstone-boson equivalence theorem~\cite{Equivalence}, in which the $Z_L$ states are replaced by their Goldstone boson counterparts $G^0$. The three neutral processes $gg\to hh,ZZ$ and $Zh$ then take the form of,respectively, the scalar ($hh$), pseudoscalar ($G^0G^0$) and mixed ($G^0h$) pair production processes. Their amplitudes have been given  in Ref.~\cite{Plehn:1996wb} and read: 
\begin{align}
\mathcal A_{ZZ} & \approx A_{G^0G^0}\propto\sum_{i=h,H}\frac{g_{itt}\,g_{iVV}M_i^2}{\shat-M_i^2+iM_i\Gamma_i}\,F_\triangle^S+F_\Box^{PP},\nonumber\\
\mathcal A_{hh}&\propto\sum_{i=h,H}\frac{g_{itt}\,v\lambda_{ihh}}{\shat-M_i^2+iM_i\Gamma_i}\,F_\triangle^S+g_{htt}^2\,F_\Box^{SS},\label{eq:amps}\\
\mathcal A_{Zh} & \approx A_{G^0 h} \propto\left[\frac{g_{Att}\,v\lambda_{AG^0h}}{\shat-M_A^2+iM_A\Gamma_A}+\frac{M_h^2\,g_{hVV}}{\shat}\right]F_\triangle^P+g_{htt}\,F_\Box^{PS}.\nonumber
\end{align}
Here $F_\triangle^{S,P}$ are the scalar and pseudoscalar top-quark triangle form factors and $F_\Box$ the box form factors \cite{Plehn:1996wb}. The one-loop integrals are evaluated with the program QCDLoop~\cite{Carrazza:2016gav}.
For the $WW$ channel, we use the $ZZ$ amplitudes, which is a good approximation and in fact, exact in the heavy-top limit that we are using here~\footnote{We include only heavy quark loops in the boxes, as we are dealing with the $V_L/$Goldstone contributions, the only ones that interfere with the (high-mass) spin-0 resonance signal. In the $m_t \to \infty$ limit, gluons couple to the Higgs doublet through  $\log(\Phi^\dagger \Phi)$ in which the Goldstones appear in the $\frac12 G^0G^0+G^+G^-$ SU(2)-symmetric combination, so that the $WW$ and $ZZ$ box contributions coincide.}.

Only part of the non-resonant amplitude can interfere with the resonance. A spin-0 state is produced by two gluons of equal helicity and decays isotropically, so after integration over the decay angles it interferes only with the angle-averaged, equal-helicity ($J=0$) part of the box diagram. This is why $F_\Box$ enters eq.~\eqref{eq:amps} averaged over the scattering angle. The remaining box contributions (opposite gluon helicities, corresponding to higher partial waves) add to the background but not to the interference. 

We have checked the  calculation in several ways. $i)$ In the limit of a very heavy top quark, the box form factors reduce to constants $F_\Box= -2/3$, $+2/3$ and $-1$ for the three processes respectively, which  one can derive independently from the low-energy effective Lagrangian. $ii)$ We find that the  high-energy amplitudes fall off as $m_t^2/\shat$, which occurs only if the large individual terms cancel correctly. $iii)$ For a heavy scalar with SM-like couplings,  the calculation reproduces the interference pattern found in the literature. 

\subsection*{3. The structure of the interference}\vspace*{-1mm}
\label{sec:structure}

\subsubsection*{3.1 The $\mathbf{VV}$ channels}\vspace*{-1mm} 

We now discuss the interference, starting with the $gg\to VV$ case.  The $h$ and $H$ couplings satisfy the sum rule $g_{htt}\,g_{hVV}+g_{Htt}\,g_{HVV}=1$, reflecting the fact that $h$ and $H$ together unitarize $VV$ scattering exactly as the single SM Higgs does. Since the non-resonant boxes do not depend on the MSSM parameters, this implies that, at LO, the full MSSM amplitude for the $gg \! \to\! VV$ process differs from the SM one by a single pole term:
\begin{equation}
\mathcal A_{\rm MSSM}=\mathcal A_{\rm SM}+g_H\,F_\triangle(\shat)\,\shat\,\bigg[\frac{1}{\shat-M_H^2+iM_H\Gamma_H}-\frac{1}{\shat-M_h^2}\bigg],\qquad g_H\equiv g_{Htt}\,g_{HVV}.
\label{eq:master}
\end{equation}
The heavy $H$ thus modifies the SM prediction near $M_H$, where it produces a resonance, and below $M_H$, where it slightly reduces the off-shell contribution of the light $h$.

 Near the resonance, any of the amplitudes in Eq.~\eqref{eq:amps} can be written as a pole term with residue $\mathcal R$ plus a smooth, non-resonant part $\mathcal{N} (\shat)$, which contains the light Higgs exchange and the box contributions. The line shape relative to the non-resonant background is then
\begin{equation}
\frac{|\mathcal A|^2}{|\mathcal{N}|^2}=\left|1+\frac{\kappa}{\epsilon+i}\right|^2 \ {\rm with} \  \epsilon\equiv\frac{\shat-M^2}{M\Gamma} \ {\rm and} \ 
\kappa\equiv\frac{\mathcal R}{M\Gamma\,\mathcal{N}(M^2)} .
\label{eq:RVV}
\end{equation}
This is a rather classic profile~\cite{Fano:1961zz}. The complex number $\kappa$ is the ratio of the resonant to the non-resonant amplitude at the peak, and it controls everything. The pure resonance, added without interference, has height $|\kappa|^2$ relative to the background. The interference distorts the line shape by an amount of relative size $1/|\kappa|$: the real part of $\kappa$ produces an antisymmetric peak--dip structure, which integrates to almost zero, while the imaginary part produces a symmetric term that changes the rate by $2\,{\rm Im}\,\kappa/|\kappa|^2$ relative to the signal. The narrow-width approximation is valid only for $|\kappa|\gg1$: {see the Appendix for details.} 

It turns out that $|\kappa|$ is small in the MSSM, and a  simple estimate explains why this special situation occurs. In the $m_t \! \to \! \infty$ limit, the non-resonant $VV$ amplitude reduces to the off-shell exchange of the light $h$ boson, and eqs.~\eqref{eq:master},\eqref{eq:RVV} give a real expression for $\kappa$:
\begin{equation}
\kappa_{VV}\simeq g_H\,(M_H^2-M_h^2)/(M_H\Gamma_H) \approx g_H\,{M_H}/{\Gamma_H}.
\label{eq:kVVsimple}
\end{equation}
For a heavy scalar with SM-like couplings, $g_H\approx1$ and $\kappa\approx M/\Gamma$, which is about $20$ for a total width of $5\%$. Interference is usually considered relevant only for broad resonances. However, in the MSSM $g_H=g_{Htt}\,g_{HVV}\simeq-\cot\beta\cos(\beta-\alpha)$ is small, while above the $t\bar t$ threshold the width is dominated by $H\to t\bar t$ with $\Gamma_H\propto \beta_t^3$. Combining the two,
\begin{equation}
\kappa_{VV}\approx 17\,|\cos(\beta-\alpha)|\tan\beta\, \beta_t^{-3},\qquad |\cos(\beta-\alpha)|\tan\beta\simeq(2\text{--}3)\,{M_Z^2}/{M_A^2}.
\end{equation}
The interference parameter therefore falls like $1/M_A^2$ and is almost independent of $\tan\beta$. It reaches values of order unity at $M_A\approx700$~GeV, where the total width is still only $1$--$2\%$ of the mass. The suppression of the coupling of $H$ to the final state, and not a large total width, is what makes the interference important in this case.
 
However, the estimate in Eq.~\eqref{eq:kVVsimple} misses an essential ingredient. In the SM, the amplitude for $gg\to Z_LZ_L$ through light $h$ exchange grows with energy and unitarity is restored by the top-quark box, with which it interferes destructively above $2m_t$. The non-resonant amplitude that interferes with $H$ is the sum of the two, and it is substantially smaller than the light-$h$ contribution alone. At $M_A=500$~GeV and $\tan\beta=2$, for example, it is reduced by a factor $\sim 2$, so $|\kappa|$ is about twice as large as Eq.~\eqref{eq:kVVsimple} suggests, and the lineshape distortion is correspondingly weaker. In addition, above $2m_t$ the box has an absorptive part, from top quarks going on shell in the loop, which gives $\kappa$ a phase. The result is a symmetric, rate-changing interference term that is absent when the top quark is treated as heavy. The same physics explains the strong cancellation between the interference of a heavy scalar with the light $h$ contributions and with the continuum found in the literature~\cite{Kauer:2015hia,ATLAS:2017spa}. A calculation with the exact $m_t$ dependence is therefore indispensable.

A final note concerns the $WW$ versus $ZZ$ cases.  Custodial symmetry implies that $h$ and $H$ couple to $WW$ and $ZZ$ with the same ratios, so the two channels have the same $\kappa$. However, they differ in how the interference appears experimentally. $ZZ\to4\ell$ reconstructs the resonance mass precisely, whereas $ZZ\to\ell\ell\nu\nu$ and $WW\to\ell\nu\ell\nu$ only measure a broad transverse mass, which averages over the line shape.\vspace*{-2mm}

\subsubsection*{3.2 The hh and Zh channels.}\vspace*{-2mm} 

The same description as above applies, with $\mathcal N$ and $\mathcal{R}$ read off from Eq.~\eqref{eq:amps}. Collecting the three channels, the interference parameters are
\begin{align}
\kappa_{VV}&= {g_H\,M_H^2\,F_\triangle^S}\, (M_H\Gamma_H)^{-1} \, \big[ g_{htt}g_{hVV}\,\dfrac{M_H^2\,F_\triangle^S}{M_H^2-M_h^2}+\left(F_\Box^{PP}-F_\triangle^S\right) \big]^{-1},\nonumber\\[4pt]
\kappa_{hh}&={g_{Htt}\,v\lambda_{Hhh}\,F_\triangle^S}\, (M_H\Gamma_H)^{-1} \, \bigg[ g_{htt}\,\dfrac{v\lambda_{hhh}\,F_\triangle^S}{M_H^2-M_h^2}+g_{htt}^2\,F_\Box^{SS}\bigg]^{-1},\label{eq:kappas}\\[4pt]
\kappa_{Zh}&={g_{Att}\,v\lambda_{AG^0h}\,F_\triangle^P}\, (M_A\Gamma_A)^{-1} \, \bigg[g_{hVV}\,\dfrac{M_h^2}{M_A^2}\,F_\triangle^P+g_{htt}\,F_\Box^{PS}\bigg]^{-1},
\nonumber
\end{align}
with all form factors evaluated at $\shat=M_H^2$ or $M_A^2$. In $\kappa_{VV}$, the first term in the denominator is the contribution of the off-shell light $h$ boson and the second is that of the top-quark box, which vanishes for a very heavy top quark. In this limit one has
\beq
F_\triangle^S\to2/3 \, , ~ F_\Box^{SS}\to-2/3 \, , ~F_\triangle^P\to1 \, , ~ F_\Box^{PS}\to-1 \, ,
\eeq
and $ \kappa_{VV}$ reduces to eq.~\eqref{eq:kVVsimple} while the other two cases become
\begin{equation}
\kappa_{hh}\to\frac{g_{Htt}\,v\lambda_{Hhh}}{M_H\Gamma_H\left[g_{htt}\,v\lambda_{hhh}/(M_H^2\!-\!M_h^2)\!-\!g_{htt}^2\right]},\
\kappa_{Zh}\to\frac{\cot\beta\cos(\beta\!-\!\alpha)\,(M_A^2\!-\!M_h^2)}{M_A\Gamma_A\left(g_{htt}\!-\!g_{hVV}M_h^2/M_A^2\right)}.~~
\label{eq:kappasHT}
\end{equation}
These expressions are useful for quick estimates; the exact form factors reduce $|\mathcal{N}|$ above $2m_t$ and add the absorptive phase discussed above.

In the $hh$ case, the non-resonant amplitude is the familiar destructive combination of the light-Higgs triangle and the box, which is negative above the $hh$ threshold (in the heavy-top limit it is proportional to $3M_h^2/(\shat-M_h^2)-1$). The resonant residue $g_{Htt}\lambda_{Hhh}$ is positive in the hMSSM at low $\tan\beta$, so that one has ${\rm Re}\,\kappa_{hh}<0$.

In the case of the $Zh$ channel, the residue is proportional to $\cot\beta\cos(\beta\!-\!\alpha)(M_h^2\!-\!M_A^2)$, and the non-resonant amplitude, the box plus the $G^0$ exchange, has the opposite sign, so again ${\rm Re}\,\kappa_{Zh}<0$. Since the $t\bar t$ width of a pseudoscalar grows like $\beta_t$ rather than $\beta_t^3$, $|\kappa_{Zh}|\approx17\,|\cos(\beta-\alpha)|\tan\beta/\beta_t$ falls faster than $|\kappa_{VV}|$ just above the threshold.

For both channels the peak--dip structure is therefore reversed relative to $VV$: the excess lies below the resonance mass and the deficit above it. The $hh$ channel has the largest $|\kappa|$ value, because $H\to hh$ is not suppressed by $\cos(\beta-\alpha)$ to the same extent.\vspace*{-1mm}

\subsubsection*{3.3 Detector resolution}\vspace*{-1mm} 

The experimental mass resolutions, a few per cent of the mass or more, are usually larger than the ratio $\Gamma/M$. The resolution therefore largely washes out the antisymmetric part of the interference, while the symmetric part survives. A simple estimate of the change of the signal yield obtained in a narrow-width analysis, valid when the resolution $\sigma_m=r\,m$ is larger than the resonance total width, is
\begin{equation}
\delta\simeq ({2\,{\rm Im}\,\kappa+c\,r\,p\,{\rm Re}\,\kappa})/{|\kappa|^2} \ {\rm with} \ c=4/\sqrt\pi\approx2.3 .
\label{eq:delta}
\end{equation}
The first term is the symmetric interference and does not depend on the resolution. The second term arises because the non-resonant spectrum falls steeply with mass, $d\sigma/dm\propto m^p$ with $p\approx-4$ to $-6$ at the LHC. The antisymmetric peak--dip structure is then weighted unequally on its two sides, and a broad template picks up a net contribution proportional to the resolution. For a resolution of $1.5\%$, typical of $ZZ\to4\ell$, $|\kappa|=3$ and $p=-4$ and the second term amounts to $-4.5\%$, while a phase of only $0.1$ in $\kappa$ gives $+6.7\%$ through the first term. For the much broader transverse-mass observables of $ZZ\!\to\! \ell\ell\nu\nu$ and $WW\!\to\!\ell\nu\ell\nu$, the two terms are comparable and can cancel. The antisymmetric part also shifts the apparent resonance mass, by the same mechanism that shifts the measured diphoton mass of the 125~GeV Higgs boson~\cite{Martin:2012xc}. In the next section we compute the effect numerically, smearing the distributions with a Gaussian resolution and fitting a narrow-width signal template; a proper and more realistic treatment is left to the experimental analyses.

\subsection*{4. Results}
\label{sec:results}

\begin{figure}[!ht]
\centering
\includegraphics[width=\textwidth]{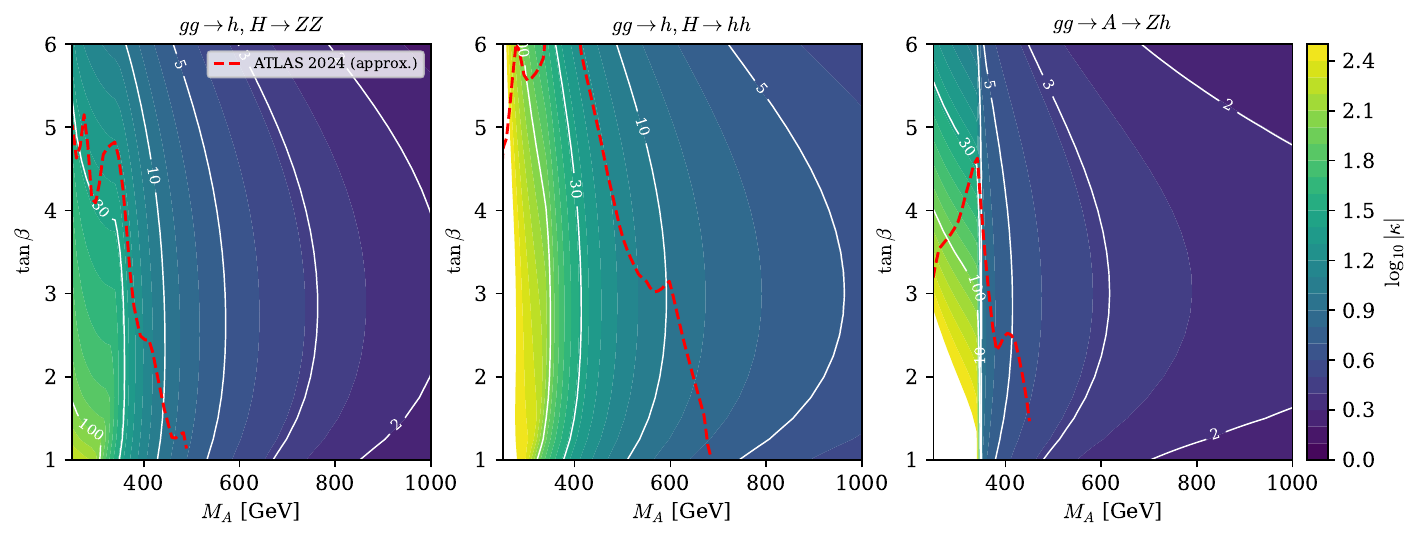}
\vspace*{-1cm}
\caption{The interference parameter $|\kappa|$ in the hMSSM for the $ZZ$ (it is similar for $WW$), $hh$ and $Zh$ channels. The interference is important when $|\kappa|$ is of order a few. The red dashed lines are the observed ATLAS exclusion boundaries of the corresponding searches, extracted from Fig.~1(c) of the 2024 search summary~\cite{ATLPUB2024008}.}
\label{fig:kappa}
\vspace*{-2mm}
\end{figure}

Let us now summarize our results. First,  Table~\ref{tab:bench} and Fig.~\ref{fig:kappa} show some values of the parameters relevant for the interference across the hMSSM $[\tan\beta, M_A]$ plane. In 
Table~\ref{tab:bench}, the Higgs masses and couplings are given with their total decay widths and the decay branching ratios,  as well as the resulting $\kappa$ parameters in all production channels. 

Below the $t\bar t$ threshold the heavy $H$ and $A$ states are extremely narrow and decay largely into $VV$, $hh$ or $Zh$ pairs, so that $|\kappa|$ is of order $10$--$10^3$ and the narrow-width approximation should be excellent. This region is nevertheless excluded by the precision measurement of the light $h$ couplings at the LHC through the signal strengths \cite{ATLPUB2024008}.

Above the threshold the $t\bar t$ decays broaden the resonances while their couplings to the final states of interest decrease, and $|\kappa|$ falls rapidly and is nearly independent of $\tan\beta$. For instance, at $\tan\beta=2$, it reaches approximately $4.5$ (for $ZZ$), $9$ (for $hh$) and $3$ (for $Zh$) at a Higgs mass $M_A=600$~GeV, $2.7$, $6$ and $2.4$ at $M_A=800$~GeV, and $2$, $4.6$ and $2$ at $M_A =1$~TeV. In all three channels, one has ${\rm Im}\,\kappa>0$ at the masses of interest, so the interference increases the signal rate. According to eq.~\eqref{eq:delta}, at $\tan\beta=2$ the increase is  $9$, $20$ and $50\%$ for $ZZ$ at $M_A=500$, $600$ and $800$~GeV, $6$, $10$ and $17\%$ for $hh$, and about $10\%$ for $Zh$ at $M_A=500$--$600$~GeV. These are large effects for resonances whose widths are only $1$--$2\%$ of their masses, so a $5\%$ $\Gamma/M$ criterion would not signal them. 

\begin{table}[!ht]
\vspace*{-2mm}
\centering\small\setlength{\tabcolsep}{4pt}
\begin{tabular}{rrrrrrrr ccc}
\toprule
$M_A$ & $t_\beta$ & $M_H$ & $c_{\beta-\alpha}$ & $\Gamma_H$ & BR$^H_{ZZ}$ & BR$^H_{hh}$ & BR$^A_{Zh}$ & $\kappa_{ZZ}$ & $\kappa_{hh}$ & $\kappa_{Zh}$\\
\midrule
350 & 3 & 357 & $-0.068$ & 0.41 & 5.7 & 50. & 3.5 & $34+0.6i$ & $-96+6.5i$ & $-10.+0.6i$\\
400 & 2 & 412 & $-0.070$ & 1.87 & 2.2 & 18 & 1.7 & $15+2.1i$ & $-32+9.7i$ & $-5.2+0.8i$\\
500 & 2 & 509 & $-0.043$ & 3.85 & 0.87 & 7.5 & 1.0 & $6.5+2.1i$ & $-13+5.9i$ & $-3.5+0.6i$\\
600 & 2 & 607 & $-0.030$ & 5.56 & 0.51 & 4.5 & 0.73 & $4.0+2.0i$ & $-8.2+4.3i$ & $-2.9+0.4i$\\
800 & 2 & 805 & $-0.016$ & 8.50 & 0.25 & 2.3 & 0.43 & $2.0+1.8i$ & $-5.1+2.9i$ & $-2.4+0.1i$\\
600 & 4 & 602 & $-0.017$ & 1.57 & 0.55 & 5.2 & 0.84 & $3.8+1.8i$ & $-8.2+4.3i$ & $-2.9+0.4i$\\
\bottomrule
\end{tabular}
\caption{hMSSM benchmark points: the masses $M_A$ and $M_H$ [in GeV], $\tan\beta$ and $\cos(\beta-\alpha)$, the width of $H$ (that of $A$ is similar except just above the $t\bar t$ threshold), the relevant branching ratios [in \%], and the interference parameters $\kappa$ of the three channels given by eq.~\eqref{eq:kappas}, calculated with the exact top-mass dependence.}
\label{tab:bench}
\vspace*{-5mm}
\end{table}

\begin{figure}[!ht]
\centering
\includegraphics[width=\textwidth]{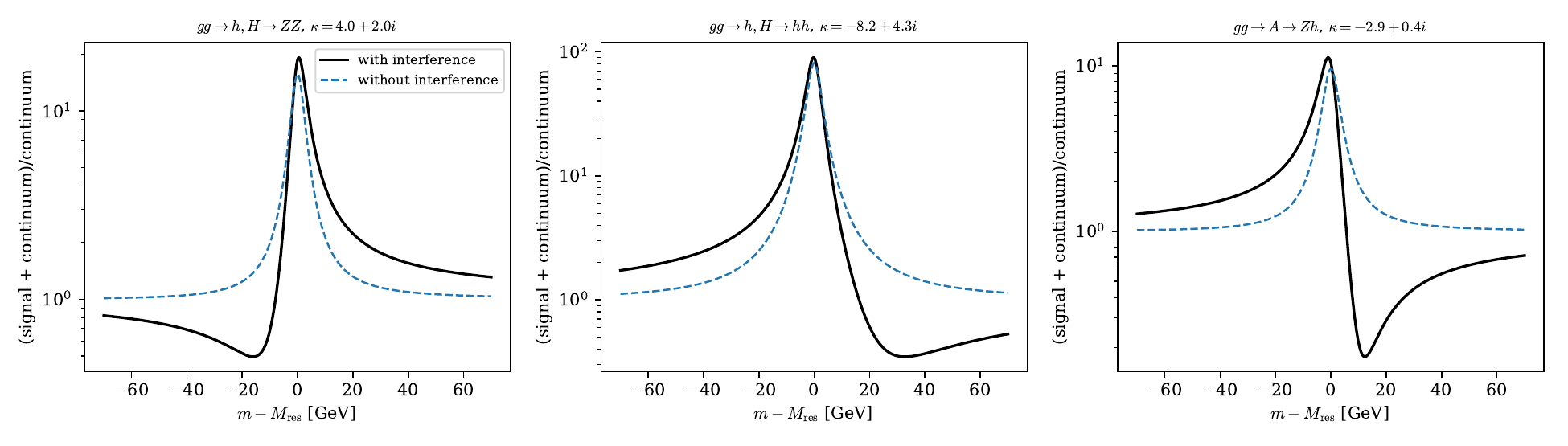}
\vspace*{-1cm}
\caption{Line shapes at $M_A=600$~GeV and $\tan\beta=2$, relative to the non-resonant top-induced continuum, with (solid) and without (dashed) interference.}
\label{fig:lineshape}
\end{figure}

Figure~\ref{fig:lineshape} shows the resulting line shapes at $M_A=600$~GeV and $\tan\beta=2$. The interference produces the expected asymmetric distortions: a deficit below and an excess above the resonance for $ZZ$, and the reverse for $hh$ and $Zh$. It also produces an overall enhancement of the resonance.

\begin{figure}[!ht]
\centering
\includegraphics[width=0.6\textwidth]{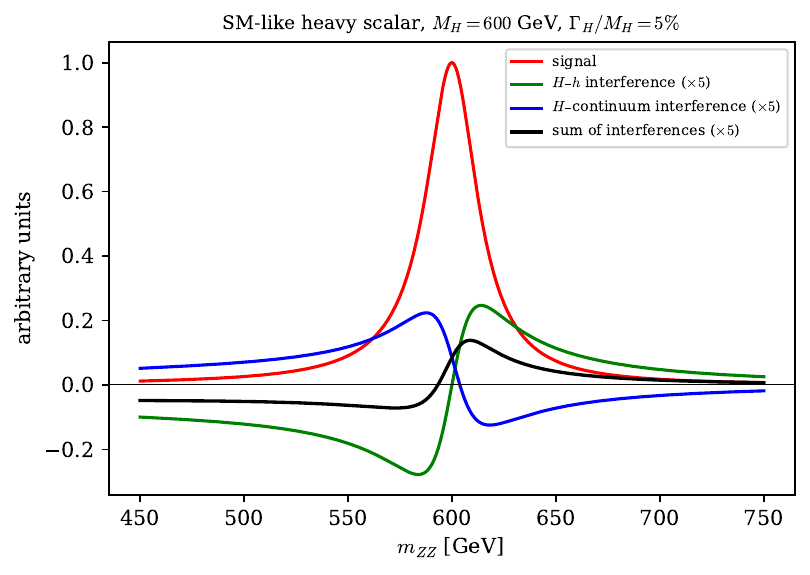}
\vspace*{-.5cm}
\caption{The cross-checked $m_{ZZ}$ distribution for a heavy scalar with SM-like couplings, $M_H=600$~GeV and $\Gamma_H/M_H=5\%$, decomposed into signal and the interferences with the off-shell light Higgs boson and with the continuum (magnified by a factor of 5).}
\label{fig:validation}
\vspace*{-3mm}
\end{figure}

We have cross-checked our result.  For a heavy scalar with SM-like couplings and $\Gamma_H/M_H=5\%$, the configuration studied by the  ATLAS collaboration~\cite{ATLAS:2017spa} and in the singlet Higgs model~\cite{Kauer:2015hia}, our calculation reproduces the known pattern, shown in Fig.~\ref{fig:validation}. The interference with the continuum is positive below and negative above the resonance, the interference with the light Higgs boson has the opposite behavior, and the two largely cancel, leaving a net enhancement of about $1\%$ within $\pm 2 \Gamma$ of the peak.

\begin{figure}[t]
\centering
\includegraphics[width=\textwidth]{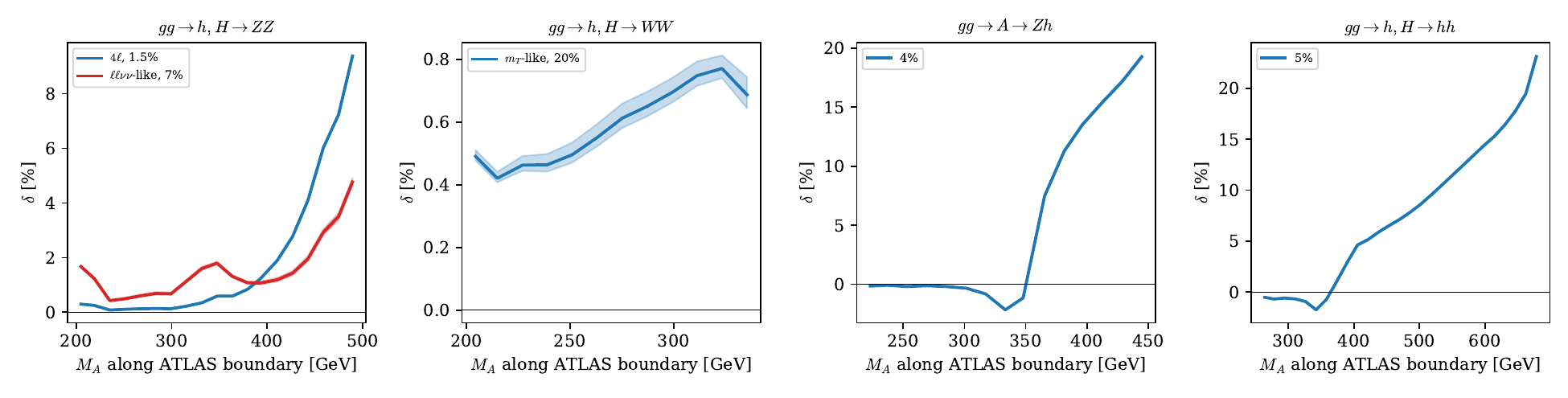}
\vspace*{-5mm}
\caption{Changes of the fitted signal strengths due to interference along the observed ATLAS exclusion boundaries~\cite{ATLPUB2024008} for $H\to ZZ$, $H\to WW$, $A\to Zh$ and $H\to hh$. For $ZZ$, the $4\ell$ (blue, $1.5\%$ resolution) and $\ell\ell\nu\nu$-like (red, $7\%$ resolution) channels are shown; for $WW$ an effective resolution of $20\%$ is used. {We use the NNPDF2.3 LO gluon luminosity function~\cite{NNPDF23} with the scale $\mu_F=M_{H/A}$, and the bands show the effects of varying the scale by a factor of two.}}
\label{fig:bound}
\end{figure}

We now discuss the impact of the interference on the present Higgs searches at the LHC. We use the most recent ATLAS summary of hMSSM constraints~\cite{ATLPUB2024008} (in particular Fig.~1c).  It contains updated $H\to WW$ (with $139$~fb$^{-1}$ luminosity)~\cite{ATLAS:WW2022} and resonant $H\to hh$~\cite{ATLAS:hh2023} results, an updated $h$-coupling fit~\cite{ATLAS:hcoup2024}, and, for the first time, the gluon-fusion $H/A\to t\bar t$ search~\cite{ATLAS:2024tt}. {The spectra are weighted with the NNPDF2.3 LO gluon luminosity function
\cite{NNPDF23}, using a factorization scale 
$\mu_F=M_{H/A}$ that we vary by a factor of two.} The data giving the ATLAS exclusion contours were extracted from Fig. 1c of Ref.~\cite{ATLPUB2024008}. 
For the experimental modeling, we use Gaussian resolutions with effective values for the transverse-mass observables, and a signal-template fit with a fixed background. 

Figure~\ref{fig:bound} shows the changes $\delta$ of the fitted signal strengths along the observed exclusion boundaries of the four channels studied here. We smear the distributions with the typical Gaussian resolutions of $1.5\%$ ($ZZ\to4\ell$), $7\%$ ($ZZ\to\ell\ell\nu\nu$), $20\%$ ($WW\to e\nu\mu\nu$), $4\%$ ($A\to Zh$) and $5\%$ ($H\to hh$). The results are as follows. 

$\bullet$ {$H\to ZZ\to4\ell$:} the change is negligible below the $t\bar t$ threshold and grows to about $+10\%$ at the end of the boundary, where $M_A\approx 500$~GeV.

$\bullet$ {$ZZ\to\ell\ell\nu\nu$ and $WW$:} the symmetric enhancement is largely compensated by the effect of the falling spectrum, leaving changes of at most $5\%$ and $1\%$, respectively.

$\bullet$ {$A\to Zh$:} $|\delta|\lesssim2\%$ below the $t\bar t$ threshold; just above it $\delta$ jumps to about $+7\%$ and grows to about $+20\%$ at the end of the boundary, where $M_A\approx450$~GeV.

 $\bullet$ {$H\to hh$:} $|\delta|\lesssim 2\%$ below the $t\bar t$ threshold; above it the signal increases from about $5\%$ at $M_A\approx400$~GeV to $12$--$23\%$ in the still relevant region $M_A\approx550$--$700$~GeV.

\begin{figure}[t]
\centering
\includegraphics[width=\textwidth]{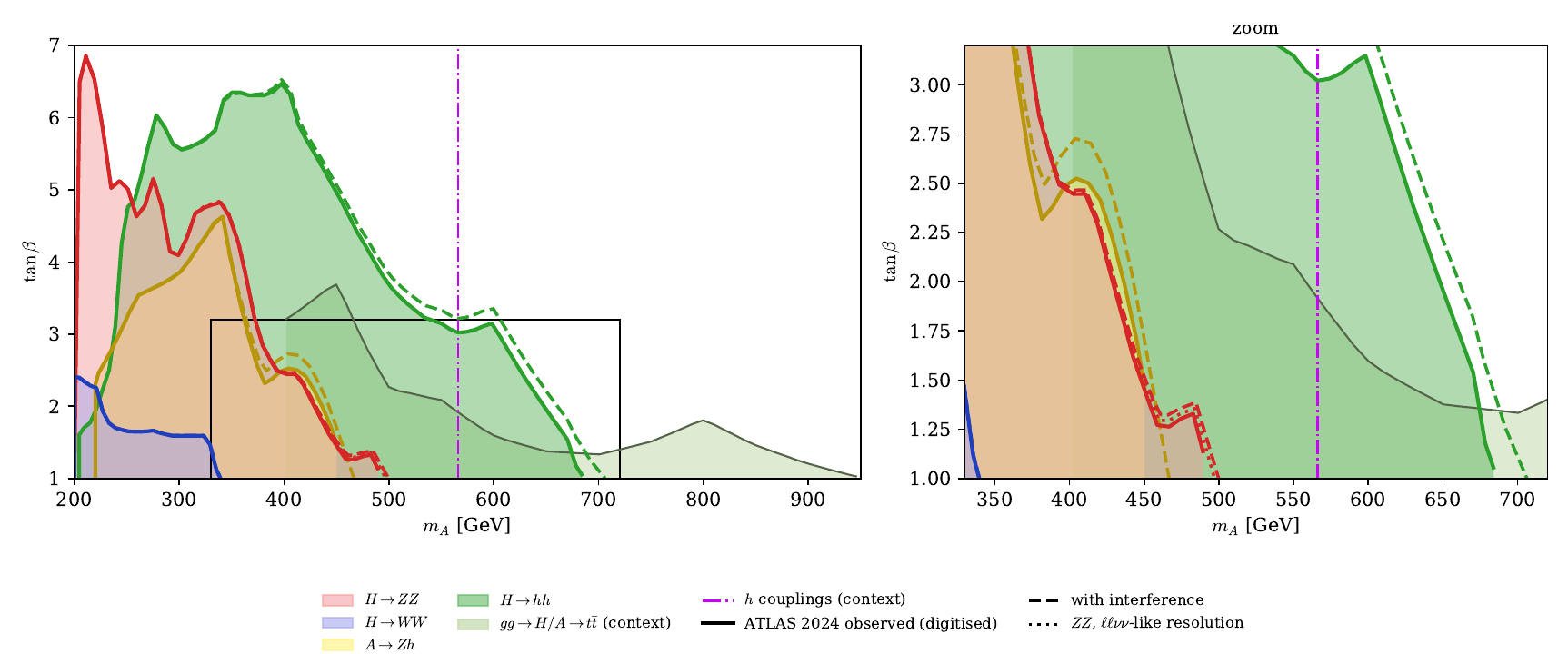}
\vspace*{-5mm}
\caption{The hMSSM exclusion regions of the channels studied here, as in Fig.~1(c) of Ref.~\cite{ATLPUB2024008} (solid lines and shading: ATLAS observed), compared with the boundaries including interference (dashed; dotted: $ZZ$ with an $\ell\ell\nu\nu$-like resolution). For context, the region excluded by $gg\to H/A\to t\bar t$ (light green) and the observed $h$-coupling limit (dash-dotted; excluded to its left) are also shown. The right panel zooms into the region of the high-mass ends of the $ZZ$, $Zh$ and $hh$ contours.}
\label{fig:compare}
\end{figure}

To translate these numbers into the exclusion contours $[M_A,\tan\beta]$ plane, we assume that each ATLAS boundary is the line where the predicted signal rate equals the observed limit. Evaluating our LO rate along the boundary then reconstructs the shape of the limit as a function of the resonance mass; the overall normalization of our rate cancels out. We then find where the rate including interference equals the same limit. The edges set by kinematic thresholds, such as $M_H=2M_h$ for $H\to hh$, are left unchanged, as is the constraint from the $h$ couplings, which involves only the on-shell light $h$ state. The result is shown in Fig.~\ref{fig:compare}. All contours move outward in the following ways: 

$\bullet$ The $ZZ$ contour moves by at most $4$~GeV in $M_A$ ($2$~GeV for the $\ell\ell\nu\nu$-like resolution); its edge at $\tan\beta=1$ moves from about $490$ to $500$~GeV;

$\bullet$ The $Zh$ contour moves by up to $\Delta\tan\beta\approx0.2$ along the flat part of its boundary at $M_A=380$--$415$~GeV, and from about $460$ to $470$~GeV at $\tan\beta=1$;

$\bullet$ The $WW$ contour does not move at all;

$\bullet$ The $hh$ contour moves by up to $15$~GeV in $M_A$ or $0.2$ in $\tan\beta$ at its high-mass end, which moves its edge at $\tan\beta=1$ from $686$ to $706$~GeV.

The updated $h$-coupling fit~\cite{ATLPUB2024008} now excludes $M_A\lesssim560$~GeV for all $\tan\beta$, which covers the whole $ZZ$, $WW$ and $Zh$ regions, and the $t\bar t$ search covers $M_A=400$--$950$~GeV for $\tan\beta\lesssim1.3$--$3.5$. The practical impact today is therefore confined to the high-mass end of the $hh$ contour, $M_A\approx560$--$710$~GeV, between the $t\bar t$ and $hh$ boundaries. The effects grow with $M_A$, where $|\kappa|$ approaches the value $1$--$3$, and will be important for the higher masses that the high-luminosity LHC will probe.

We note finally that eq.~\eqref{eq:master} also shows that, below $M_H$, the heavy state reduces the off-shell contribution of the light Higgs boson to $gg\to VV$, whose coupling factor is $g_{htt}g_{hVV}=1-g_H$. For $M_A=400$--$500$~GeV and $\tan\beta=1.5$--$3$ the reduction of the amplitude is $1$--$6\%$. This is too small for the present off-shell measurements, but it is a qualitatively new, indirect probe of the heavy CP-even state.

\subsection*{5. Conclusions}
\label{sec:conclusions}

The searches for the heavy neutral MSSM Higgs bosons decaying into $ZZ$, $WW$, $hh$ pairs and $Zh$ final states are usually interpreted assuming narrow resonances that do not interfere with the SM background. We have shown that at low values of $\tan\beta$, this assumption can fail even though the resonances are narrow, with total decay widths of only a few per cent of their masses. The reason is that the heavy $H/A$ bosons couple only weakly to these final states, so that the resonant amplitude is not much larger than the SM background amplitude it interferes with. The size of the effect is set by a single complex number per channel, $\kappa$, the ratio of the two amplitudes at the peak, which we have calculated at leading order in perturbation theory (as we are calculating only the relative effect, higher-order corrections should not be too relevant) but with the exact top-quark mass dependence, which plays an important role. It should replace the resonance total width as the criterion for the validity of the narrow-width interpretations.

In the context of the hMSSM~\footnote{This analysis can be readily extended to another MSSM scheme for including the radiative corrections and even to a general 2HDM by simply adjusting the masses and couplings of the $H$ and $A$ states.}, we have discussed the impact of the interference with the background for the limits set by the ATLAS collaboration in its various MSSM Higgs searches at the LHC up to 2024. It turns out that the consequences on these limits are modest. Above the  $t\bar t$ kinematical threshold, the interference increases the signal rate by up to tens of per cent and distorts its line shape.  The ATLAS $ZZ$ and $WW$ exclusion contours move by at most a few GeV and the $Zh$ contour by up to $0.2$ in $\tan\beta$, all in a region now excluded by the light $h$-coupling measurements. The high-mass end of the $hh$ exclusion contour, which is still relevant, extends by up to $15$--$20$~GeV. For the masses of $600$~GeV to $1$~TeV that the high-luminosity LHC will probe, however, interference becomes an $\mathcal O(1)$ effect in all three channels, and resonance line shapes including interference should replace the narrow-width signal templates.

We have used  various approximations and simplifications in our analysis. For instance, on the theory side the gauge bosons were treated using the equivalence theorem, which is accurate up to ${\cal O}(M_Z^2/\shat)$ corrections i.e. a few percent for masses of about $500$~GeV, the total Higgs widths have been calculated at NLO only, which might generate a few percent error, and  the trilinear Higgs couplings have been treated using the approximate hMSSM prescription. On the experimental side, we use a Gaussian resolution and a signal fit assuming a fixed  background, with the ATLAS experimental contours extracted from a figure. None of these approximations should affect the main qualitative conclusions: that the interference is controlled by $\kappa$, that $|\kappa|$ becomes of order a few above the $t\bar t$ threshold, and that the absorptive part increases the signal rate. Nevertheless, a complete treatment within the experimental analyses would be needed for a quantitative reinterpretation of the exclusion limits.\bigskip

\noindent {\bf Acknowledgements}: 
We thank 
Claude Anthropic for valuable technical help. A.D. is supported by the  grant PID2024-161668NB-I00 and he thanks  CERN for hospitality.

\subsection*{Appendix: The Fano profile for interference}
\setcounter{equation}{0}\renewcommand{\theequation}{A.\arabic{equation}}
\label{app:fano}

In this Appendix we describe how the size of the interference is controlled by $\kappa$ and how $\kappa$ depends on the total width and on the couplings of the resonance.

Near the resonance, the amplitude for one of the processes that we are considering is the sum of a smooth, non-resonant part ${\cal N}$ and a pole term,
\begin{equation}
A(\hat s)={\cal N}+\frac{\cal R}{\hat s-M^2+iM\Gamma} \ \Rightarrow\ \  \frac{A}{\cal N}=1+\frac{\kappa}{\epsilon+i},\ \  \epsilon=\frac{\hat s-M^2}{M\Gamma},\ \  \kappa=\frac{\cal R}{M\Gamma\,{\cal N}}.
\label{eq:A1}
\end{equation}
When squaring, one gets three terms, the background, the signal and the interference:
\begin{equation}
{|A|^2}/{|{\cal N}|^2}=1+{|\kappa|^2}/{(\epsilon^2+1)} + ({2\,\epsilon\,{\rm Re}\,\kappa+2\,{\rm Im}\,\kappa})/({\epsilon^2+1}).
\label{eq:A2}
\end{equation}
The signal is a Breit--Wigner peak of height $|\kappa|^2$ in units of the background. The interference is of order $|\kappa|$, so it is of order $1/|\kappa|$ relative to the signal: it is negligible for $|\kappa|\gg1$ and comparable to the signal for $|\kappa|\sim1$. Its part proportional to ${\rm Re}\,\kappa$ is antisymmetric in $\epsilon$, with a dip on one side of the peak and an excess on the other, and integrates to zero. On the other hand, its part proportional to ${\rm Im}\,\kappa$ has the same Breit--Wigner shape as the signal. Integrating over $\epsilon$, the signal gives $\pi|\kappa|^2$ and the interference $2\pi\,{\rm Im}\,\kappa$, so that interference changes the rate by $\delta={2\,{\rm Im}\,\kappa}/{|\kappa|^2} $ relative to the signal. This is the first term in Eq.~(\ref{eq:delta}); the second one arises only when a falling spectrum and a finite resolution weight the two sides of the antisymmetric term differently.

 Since $|\kappa|^2$ is the ratio of the signal to the background at the peak, it can be read off from the Breit--Wigner formula. At $\hat s=M^2$, the resonant partonic cross section depends on the couplings and on the width only through the decay branching ratios (BRs):
\begin{equation}
\hat\sigma^{gg\to\Phi\to XX}_{\hat s=M^2}\propto \frac{{\rm BR}(\Phi\to gg)\,{\rm BR}(\Phi\to F)}{M^2}  
\Rightarrow |\kappa|^2\propto\frac{{\rm BR}(\Phi\to gg)\,{\rm BR}(\Phi\to F)}{M^2\,\hat\sigma_B^{J=0}(M^2)},
\label{eq:A4}
\end{equation}
where $\hat\sigma_B^{J=0}$ is the part of the background that can interfere (equal gluon helicities, $J=0$). Hence $|\kappa|\propto g_{\rm prod}\,g_{\rm dec}/\Gamma$. The width and the couplings thus enter in a definite combination:

$\bullet$ At fixed couplings to the initial and final states, a larger width from other decay channels lowers the BRs, hence $|\kappa|$, and makes the interference relatively more important;

$\bullet$ At fixed width, a smaller coupling to the final state lowers $|\kappa|$ in the same way.

A small width is thus a guarantee that interference is negligible only if the branching ratios are large. For a heavy scalar with SM-like couplings, ${\rm BR}(H\to VV)\approx1$ and $\kappa\approx M/\Gamma$, i.e.\ $\kappa\approx20$ for $\Gamma/M=5\%$, and the interference is a few per cent of the signal. In the MSSM at $M_A\!=\!600\, $GeV and $\tan\beta\!=\!2$, $\Gamma/M \simeq 1\%$ but ${\rm BR}(H\to ZZ)\approx0.5\%$ as $H$ decays mostly into $t\bar t$. Then $\kappa_{ZZ}=4+2i$ and one gets a $20\%$ increase of the rate.

\begin{figure}[h]
\centering
\includegraphics[width=\textwidth]{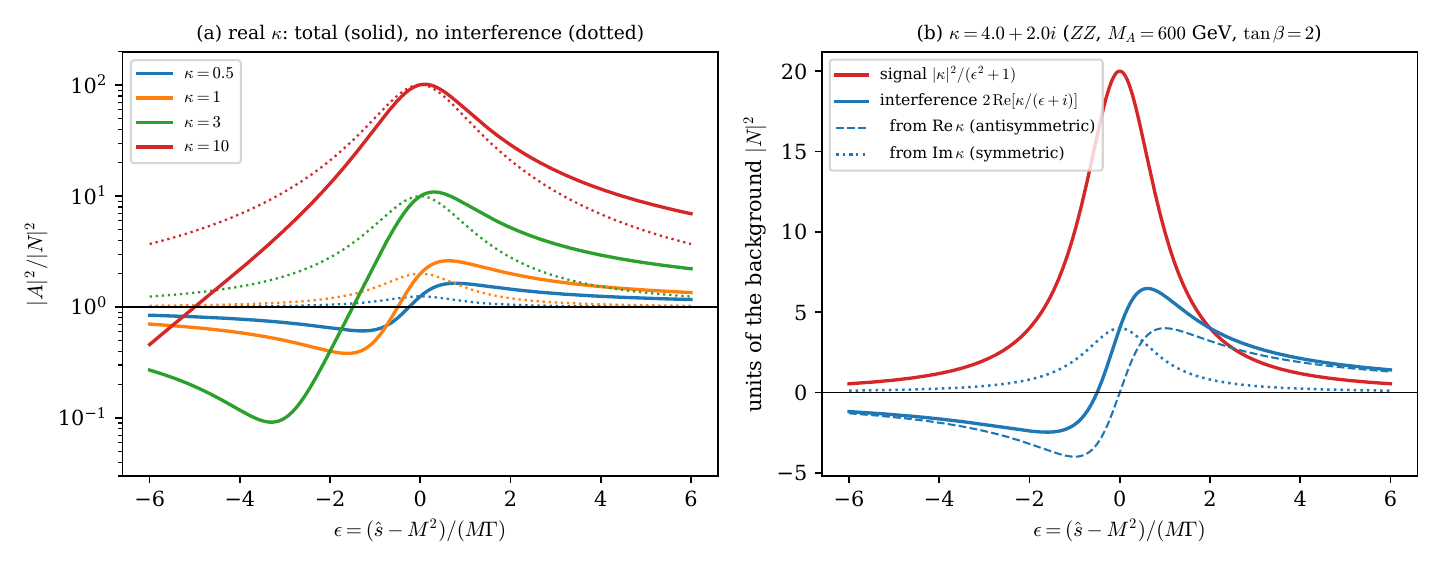}
\caption{(a) Fano profile, eq.~(\ref{eq:A2}), for real $\kappa$ (solid), compared with the resonance added without interference (dotted). (b) Decomposition of the profile for $\kappa=4+2i$, the value for $ZZ$ at $M_A=600$~GeV and $\tan\beta=2$ (Table~\ref{tab:bench}), into signal and interference, with the interference split into its antisymmetric (${\rm Re}\,\kappa$) and symmetric (${\rm Im}\,\kappa$) parts.}
\label{fig:fanoapp}
\end{figure}

Figure~\ref{fig:fanoapp} illustrates these statements. The left panel shows the profile (\ref{eq:A2}) for real $\kappa$: for $\kappa=10$ the resonance is close to the incoherent sum of signal and background, while for $\kappa\lesssim3$ the line shape is strongly distorted. The right panel decomposes the profile for $\kappa=4+2i$: the antisymmetric interference term from ${\rm Re}\,\kappa$ integrates to zero, and the symmetric term from ${\rm Im}\,\kappa$, $20\%$ of the signal, changes the rate.

\end{document}